\documentclass[11pt,superscriptaddress,reprint, preprintnumbers,nofootinbib,prd]{revtex4-1}

\usepackage[utf8]{inputenc}
\usepackage{amsmath}
\usepackage{graphicx}
\usepackage{amssymb}
\usepackage{bbold}
\usepackage{physics}
\usepackage{comment}
\usepackage[normalem]{ulem}  
\usepackage{xcolor}
\usepackage[breaklinks]{hyperref}
\hypersetup{colorlinks=true,allcolors=[rgb]{0.5,0.0,0.5}}

\newcommand{\orcid}[1]{\href{https://orcid.org/#1}{#1}}

\newcommand{\beq}{\begin{equation}}
\newcommand{\eeq}{\end{equation}}
\newcommand{\bea}{\begin{eqnarray}}
\newcommand{\eea}{\end{eqnarray}}

\begin{document}

\title{Monophotons at Neutrino Experiments from Neutrino Polarizability}

\author{Julia Gehrlein}

\email{julia.gehrlein@colostate.edu}
\thanks{\orcid{0000-0002-1235-0505}}
\affiliation{Physics Department, Colorado State University, Fort Collins, CO 80523, USA}

\author{Ian M. Shoemaker}
\email{shoemaker@vt.edu}
\thanks{\orcid{0000-0001-5434-3744}}
\affiliation{Center for Neutrino Physics, Department of Physics, Virginia Tech, Blacksburg, Virginia 24601, USA}

\author{Anil Thapa}

\email{a.thapa@colostate.edu}
\thanks{\orcid{0000-0003-4471-2336}}
\affiliation{Physics Department, Colorado State University, Fort Collins, CO 80523, USA}

\date{\today}

\begin{abstract}
Nontrivial electromagnetic properties of neutrinos are an avenue to physics beyond the Standard Model. To this end, we investigate the power of monophoton signals at neutrino experiments to probe a higher-dimensional operator connecting neutrinos to  SM photons dubbed, {\it neutrino polarizability}. A simplified scenario giving rise to this operator involves a new pseudo-scalar that couples to both neutrinos and photons, with clear implications for axion-like particle (ALP) and Majoron physics. By analyzing the photon energy spectrum and angular distributions, we find that NOMAD and MiniBooNE currently set the most stringent limits, while SBND and the DUNE near detector will soon provide significantly improved constraints.  

\end{abstract}

\date{\today}

\maketitle

\maketitle
\section{Introduction}

The wealth of data in the upcoming experimental neutrino program allows us to test the neutrino sector with great precision. The oscillation program with future accelerator long-baseline experiments DUNE \cite{DUNE:2020ypp} and Hyper-Kamiokande \cite{Hyper-Kamiokande:2018ofw,Hyper-Kamiokande:2025fci}, in combination with the medium-baseline reactor neutrino experiment JUNO \cite{JUNO:2015zny} and atmospheric neutrino experiments IceCube-Gen2 \cite{Ishihara:2019aao} and KM3NeT \cite{KM3Net:2016zxf}, aims to measure the remaining unknown parameters in the neutrino sector -- such as the CP-violating phase $\delta$, the neutrino mass ordering, and the octant of $\theta_{23}$ -- while improving the precision on all known oscillation parameters \cite{Denton:2022een}. Although the primary goal of these experiments is to test the standard three-flavor oscillation framework, they also offer the potential to uncover yet unknown physics in the neutrino sector such as new neutrino interactions, additional neutrino generations, or their non-trivial electromagnetic properties. These efforts are further aided by the Short-Baseline Neutrino (SBN) program at Fermilab \cite{Machado:2019oxb} and the DUNE near detector \cite{DUNE:2021tad}, both of which are promising for discovering new physics due to their large neutrino fluxes. In particular, future neutrino detectors based on the liquid argon time projection chamber (LArTPC) technology are especially appealing in probing new physics due to their excellent energy and timing resolution.

The electromagnetic properties of neutrinos have been a subject of interest since the earliest days of neutrino physics. In fact, Wolfgang Pauli's original letter~\cite{Pauli:83282} in which he proposed the neutrino to explain continuous beta decay spectra also suggested the possibility of detecting neutrinos directly via their magnetic dipole moment. Given the fact that neutrinos have masses, it is now established that the neutrino sector participates in some new physics beyond the Standard Model (BSM). As such, it behooves us to verify other neutrino properties, such as their electromagnetic interactions, to match the SM expectation. 

In the SM, neutrinos are among the most challenging particles to detect because they do not couple directly to the photon or the strong nuclear force. However, there are a number of well-motivated BSM scenarios in which neutrinos may have new, sizable interactions with the electromagnetic force (for a recent review see Ref.~\cite{Giunti:2014ixa,Giunti:2024gec}). The most studied possibilities for new neutrino electromagnetic properties include milli-charges~\cite{Foot:1989fh,Babu:1989tq,Barbiellini:1987zz,Khan:2022akj,Jana:2025eii}, magnetic dipole moments~\cite{Lee:1977tib,Petcov:1976ff,Pal:1981rm,Fujikawa:1980yx,Brdar:2020quo,Shrock:1982sc,Dvornikov:2003js,Magill:2018jla}, and charge radii~\cite{Bernabeu:2000hf,Bernabeu:2002nw,Bernabeu:2002pd}.

In this work, we investigate a less studied possible coupling of neutrinos to photons, namely the {\it neutrino polarizability}~\cite{Bansal:2022zpi}. The lowest-dimensional operator that couples neutrinos to two photons is through a Rayleigh operator which arises at dimension-8 (dimension-9) for Dirac (Majorana) neutrinos. After electroweak symmetry breaking, the effective dimension-7 operator can be written as
\begin{align}
    {\cal L}_{\rm EFT} = 
     \frac{\alpha}{8 \pi} \frac{C^{7}_{ij}}{\Lambda^3}(\bar{\nu}_i P_L\nu_j)F^{\mu\nu}\widetilde{F}_{\mu\nu} \, ,
    \label{eq:eftpol}
\end{align}
where $\widetilde{F}_{\mu \nu} = \frac{1}{2} \epsilon_{\mu \nu \rho \sigma} F^{\rho \sigma}$ is the dual electromagnetic field strength tensor, $C_{ij}^7$ is a Wilson coefficient, and $\Lambda$ is the scale of new physics. A similar dimension-7 operator also exists with $F^{\mu\nu}F_{\mu\nu}$.  
The operator of Eq.~\eqref{eq:eftpol} can be generated by introducing a light pseudo-scalar mediator that couples to both neutrinos and photons at  tree-level~\cite{Bansal:2022zpi}. Such a mediator can be naturally identified as a pseudo-Goldstone boson (pGB) of a spontaneously broken global $U(1)$ symmetry, such as a Majoron \cite{Gelmini:1980re,Chikashige:1980ui,Schechter:1981cv} or an axion-like particle (ALP) \cite{Peccei:1977hh,Peccei:1977ur,Weinberg:1977ma,Wilczek:1977pj,Arkani-Hamed:1998wff,Dienes:1999gw,Chang:1999si,DiLella:2000dn,Davidson:1981zd,Wilczek:1982rv,Ema:2016ops,Calibbi:2016hwq,Abbott:1982af,Dine:1982ah,Preskill:1982cy,Cicoli:2013ana}.
In this manuscript, we remain agnostic about the underlying ultra-violet (UV) completion and focus on deriving new constraints on this operator from neutral current neutrino scattering processes with a single photon in the final state (NC$1\gamma$) via the process $\nu\cal{N}\to\nu \cal{N}+\gamma$, where $\cal N$ represents the nucleus/nucleon. Related processes have also previously been investigated in Ref.~\cite{Bansal:2022zpi,Dutta:2025fgz}. Note that in the SM this NC$1\gamma$ final state can arise from the de-excitation of a nuclear resonance in NC scattering \cite{Wang:2013wva} and it leads to an irreducible background in electron-(anti)neutrino appearance measurements at T2K \cite{Wang:2015ivq}. 

Here we study the NC$1\gamma$ process across a variety of current and future neutrino experiments by analyzing the photon energy spectrum and angular distribution. In fact, several experiments have already searched for this final state. Notably, MicroBooNE \cite{MicroBooNE:2015bmn} recently reported a search for NC$1\gamma$ events and observed a mild excess with the local significance of approximately 2$\sigma$ \cite{MicroBooNE:2025ntu}.  In addition, previous neutrino experiments such as the Neutrino Oscillation MAgnetic Detector (NOMAD) \cite{NOMAD:2011gyy} and the T2K near detector \cite{T2K:2019odo} have also provided results on this final state.
Finally, this NC$1\gamma$ process have been  proposed as a potential explanation for the MiniBooNE low-energy excess \cite{Abdullahi:2023ejc,Dutta:2025fgz}. 

This manuscript is organized as follows: 
We describe the model assumptions and setup in Sec.~\ref{sec:model} and present our 
 analysis and experimental details in  \ref{sec:analysis}. We compare and contrast our results with relevant constraints from Appendix~\ref{sec:constraints} in Sec.~\ref{sec:results}   before we conclude in Sec.~\ref{sec:summary}.

\section{Model}
\label{sec:model}
\begin{figure}
    \centering
    \hspace{1mm}\includegraphics[width=0.45\linewidth]{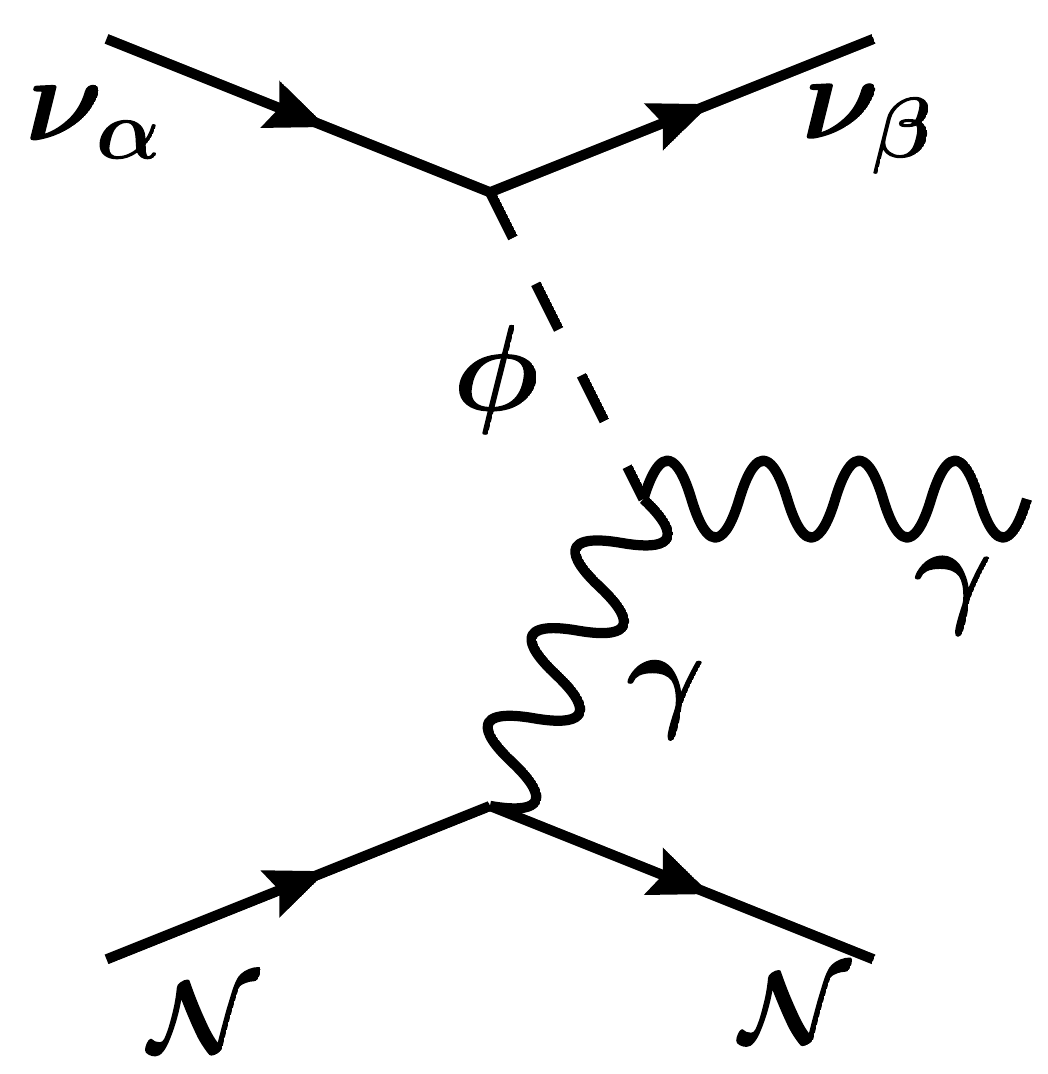}
    \caption{Tree-level Feynman diagram of neutrino scattering with a monophoton signal via a neutrino polarizability operator realized through the exchange of the pseudo-scalar $\phi$.  }
    \label{fig:treepol}
\end{figure}

A possible realization of the effective operator in Eq.~\eqref{eq:eftpol} involves the introduction of a light pseudo-scalar field $\phi$ that couples to both neutrinos and photons through the interactions 
\begin{align}
\mathcal{L}\supset 
\frac{\alpha}{8\pi}\frac{c_\gamma^\prime}{f_\phi}\phi F^{\mu\nu}\widetilde{F}_{\mu\nu}
+\frac{1}{2}c_\nu^{ij}(\overline{\nu}_iP_L\nu_j) \phi+\text{h.c.} ,
\label{eq:fulllag}
\end{align}
where $c'_\gamma$ and $c_\nu$ are dimensionless couplings, and $f_\phi$ is a dimensionful parameter associated with the UV scale (e.g., the decay constant of a pseudo-Goldstone boson).  
The tree-level exchange of $\phi$ mediates the neutrino-photon scattering, which leads to the NC$1\gamma$ final state, is shown in Fig.~\ref{fig:treepol}.
If the scalar $\phi$ is much heavier than energy and momentum exchange ($q$) in this process, it can be integrated out. This then leads to the effective operator of Eq.~\eqref{eq:eftpol}, with the matching condition 
\begin{equation}
    \frac{C^{7}}{\Lambda^3} = \frac{c_\nu c^{\prime}_\gamma}{m_\phi^2 f_\phi} \, .  
\end{equation}
For convenience, we define $g_{\phi\gamma} = \alpha c'_\gamma/(2\pi f_\phi)$. The cross-section of the process $\nu \cal{N}\to \nu \cal{N}+\gamma$ thus depends on the product of couplings $c_\nu g_{\phi\gamma}$ and the mass of the mediator, $m_\phi$. In sec.~\ref{sec:results} we will therefore derive constraints on the $(m_\phi,c_\nu g_{\phi\gamma})$ parameter space. 

At tree-level, the coupling of a pGB to fermions is generally proportional to the fermion masses \cite{Bauer:2020jbp,Chala:2020wvs}, which renders the neutrino-$\phi$ coupling negligible. However, this suppression can be overcome by considering one-loop radiative corrections~\cite{Bonilla:2021ufe}.
Majoron-neutrino couplings have been studied up to two-loops order in Ref.~\cite{Heeck:2019guh} and ALP-neutrino couplings at one-loop in Ref.~\cite{Bonilla:2023dtf}. In what follows, we remain agnostic about the specific UV complete model and focus instead on the phenomenology of the effective interaction. 
Additionally, we assume that the field $\phi$ couples to one neutrino flavor at-a-time.  The neutrino flux at the experiments we consider is dominated by muon neutrinos and to a lesser extend, electron neutrinos. In the following we will derive the constraints on $c_\nu^\mu g_{\phi\gamma}$ and $c_\nu^e g_{\phi\gamma}$.  We further assume that the resulting cross-section  is the same for neutrino or anti-neutrino final states. 
Finally, we do not rely on whether the underlying interactions are lepton number conserving or violating, making the analysis applicable to both scenarios.  

Before presenting the novel bounds derived from neutrino experiments, we briefly discuss the most relevant existing constraints on the combination of couplings $c_\nu g_{\phi\gamma}$ and the mass of the mediator $m_\phi$. Cosmological observations, particularly BBN and CMB measurements, place a lower bound on the mass of $\phi$, $m_\phi\gtrsim 5$ MeV, with a precise limit depending on the combined data sets~\cite{Sabti:2021reh}. The most stringent limit on this model from terrestrial experiments -- aside from those derived in this work -- comes from the XENONnT experiment~\cite{XENON:2022ltv}, where the light mediator leads to an enhanced rate of solar neutrino-electron scattering \cite{Bansal:2022zpi}. In the heavy mediator regime, $m_\phi^2 \gg q^2$, the constraint reads $c_{\nu}^\mu (g_{\phi \gamma} \times {\rm GeV}) \lesssim 2.32 \times 10^{-6} \left(m_\phi/{\rm MeV}\right)^2$ and $c_{\nu}^e (g_{\phi \gamma} \times {\rm GeV}) \lesssim 6.97 \times 10^{-7} \left(m_\phi/{\rm MeV}\right)^2$.  In the light mediator limit, $m_\phi^2 \gg q^2$, the bound becomes $ c_{\nu}^\mu (g_{\phi \gamma} \times {\rm GeV}) \lesssim 8.9 \times 10^{-8}$ and $ c_{\nu}^e (g_{\phi \gamma} \times {\rm GeV}) \lesssim 2.6 \times 10^{-8}$~\cite{Bansal:2022zpi}. The Borexino experiment~\cite{Borexino:2017fbd} is sensitive to the same solar neutrino-electron scattering but yields slightly weaker bounds. Additional constraints from stellar cooling, meson decays, and collider searches are also relevant; however, they typically constrain only individual couplings (e.g., $g_{\phi\gamma}$ or $c_\nu$) rather than their product. A more detail discussion of these complementary bounds is provided in Appendix~\ref{sec:constraints}.

\section{Bounds from neutrino experiments}
\label{sec:analysis}
We now turn to the discussion of different experimental probes of the monophoton signal predicted by the proposed model. We use the current existing experimental data from MiniBooNE \cite{MiniBooNE:2020pnu,MiniBooNE:2018esg}, MicroBooNE \cite{MicroBooNE:2025ntu}, T2K \cite{T2K:2019odo}, and NOMAD \cite{NOMAD:2011gyy} to constrain the model parameter space and additionally consider projected sensitivities of experiments where dedicated searches are not available yet like at the DUNE near detector \cite{DUNE:2020ypp}, SBND \cite{SBND:2025lha}, and ICARUS \cite{ICARUS:2023gpo}.

We first outline the methodology used in our analysis. The incoming neutrino energies vary significantly across different experiments --  ranging from the MeV scale to the multi-GeV -- necessitating a treatment of the scattering process in different kinematic regimes: coherent, incoherent, and deep inelastic scattering (DIS). We classify these regimes based on the momentum transfer in the $2\to3$ process, $q^2 = - (p_{\cal N}^{\rm out} - p_{\cal N}^{\rm in})^2$. Specifically, we define the coherent regime as $q^2 < 0.1\ {\rm GeV}^2$, the incoherent regime as $0.1\ {\rm GeV}^2 < q^2 < 1.8\ {\rm GeV}^2$, and the DIS regime as $q^2 > 1.8\ {\rm GeV}^2$. For the coherent and incoherent regimes, we use the appropriate nuclear form factors to account for the structure of the target nucleus or nucleon. See the Appendix~\ref{sec:details_xsec} for details on the implementation of the scattering cross-section in the different regimes.
The predicted photon spectrum of the scattering process $\nu\ {\cal N} \to \nu\ {\cal N} + \gamma$ is computed as in~ \cite{Bansal:2022zpi}
\begin{equation}
\frac{d N}{d E_\gamma}=N_p\ {\rm POT} \int d E_\nu \epsilon\left(E_\gamma\right) \phi_{\mathrm{tot}}\left(E_\nu\right) \frac{d^2 \sigma}{d E_\gamma d E_\nu},
\end{equation}
where $N_p$ is the total number of target protons in the detector, POT is the number of protons on target, $\phi_{\rm tot}( E_\nu)$ is  incoming muon or electron neutrino flux, and $\epsilon (E_\gamma)$ is the photon detection efficiency. 
The differential cross-section $d^2 \sigma/(d E_\gamma d E_\nu)$ is obtained from {\tt Madgraph} \cite{Alwall:2011uj}. We implement our model file in the {\tt FeynRules} package \cite{Christensen:2008py} and compute the cross-sections at the parton level for all regimes using the {\tt MadGraph5} event generator \cite{Alwall:2014hca}. 
Our results  are generally in agreement with those of Ref.~\cite{Bansal:2022zpi}. However,  at lower mediator masses we find constraints which are up to an order of magnitude  stronger compared to 
\cite{Bansal:2022zpi}.

For experiments with existing data -- MiniBooNE, MicroBooNE, NOMAD, T2K near detector -- we perform a chi-squared ($\chi^2$) analysis using the number of observed and expected background events, taking into account both statistical and systematic uncertainties for each experiment. For convenience, these numbers are listed in table~\ref{tab:num_events}. For MiniBooNE, we treat the neutrino and the anti-neutrino runs independently, compute the individual $\chi^2$ contribution, and then obtain the constraint on the parameter space by minimizing the sum of the respective chi-squared $\chi^2_\nu + \chi^2_{\bar \nu}$. All of these experiments provide both the photon energy and angular spectrum. Since the correlation between the two observables is not publicly available, we derive the independent constraints from each observable. We find that the photon energy spectrum provides a slightly stronger bound, and adopt it for our main results. 
\begin{table*}[!t]
    \centering
    \begin{tabular}{|c|c|c|}
   \hline
       Experiment & Observed Events& Background Events\\&& and Uncertainty \\ 
        \hline \hline
 MiniBoone ($\nu$) \cite{MiniBooNE:2020pnu}&2870 &2309.4$\pm$  48.1(stat.) $\pm$ 109.5(syst.)\\
 MiniBooNE ($\bar\nu$) \cite{MiniBooNE:2018esg}&478&398.7$\pm$ 20 (stat) $\pm$ 28.6 (syst)\\
 MicroBooNE ($\nu$) \cite{MicroBooNE:2025ntu}&678  &564 $\pm$ 24 (stat.) $\pm$ 51(syst.)\\
 NOMAD \cite{NOMAD:2011gyy}&155&129.2 $\pm$ 8.5 (stat.) $\pm$ 3.3 (syst.)\\
 T2K \cite{T2K:2019odo}&39&42 $\pm$ 5.88 (stat.)(+11.34-6.3) (syst.)\\
\hline
    \end{tabular}
     \caption{Number of observed and background events including their uncertainty for the experiments we study.   }
\label{tab:num_events}
\end{table*}

To derive forecasted constraints from upcoming experiments like DUNE, SBND, and ICARUS, we 
assume that no excess events are observed and impose the condition that the total number of predicted signal events remain below 2.7 at 90\% C.L assuming Poisson statistics.
We used published data for the neutrino fluxes, number of protons-on-target, and photon detection efficiencies for all existing and upcoming experiments; see appendices~ \ref{sec:details_analysis} and \ref{sec:exp_details} for more details on the experimental simulation and the analysis. 

\section{Results and Experimental sensitivities}
\label{sec:results}
\begin{figure}[!t]
    \centering
    \includegraphics[width=0.49\textwidth]{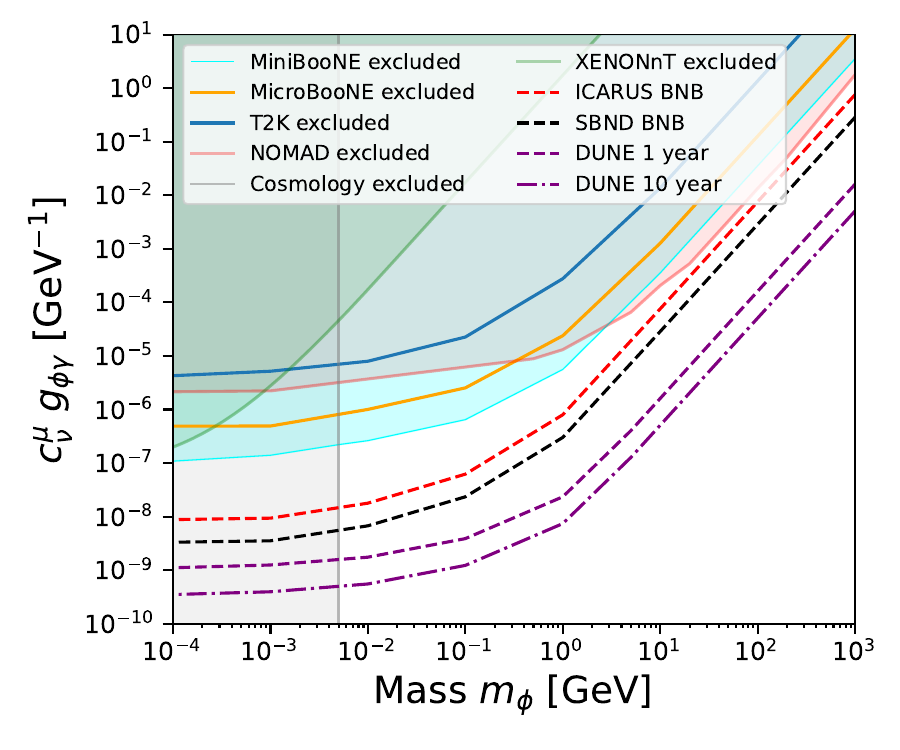}\\
    \includegraphics[width=0.49\textwidth]{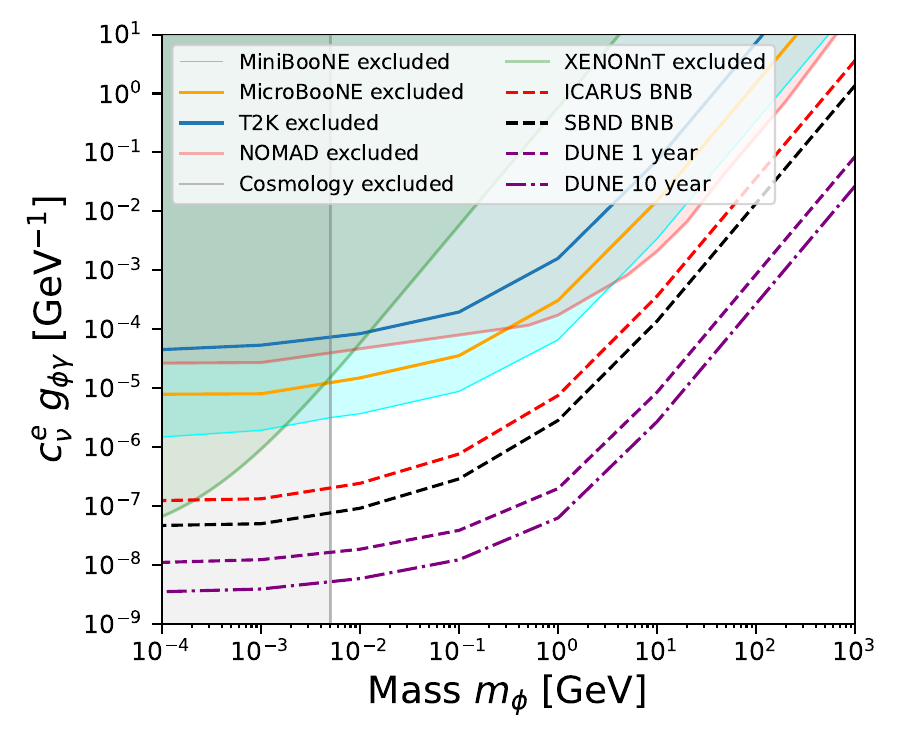}
    \caption{Exclusion region and future sensitivity from different neutrino experiments in the plane of the mass of the mediator and the product of its couplings to muon neutrinos and photons, $m_\phi$ -- ($c_\nu^\mu ~g_{\phi\gamma}$) (upper plot) and $m_\phi$ -- ($c_\nu^e ~g_{\phi\gamma}$) (lower plot). Future experimental projections are shown as dashed lines. We also show the constraints from $\nu e\to \nu e$ scattering from XENONnT in green and the bound from cosmology on neutrinophilic scalars in gray \cite{Bansal:2022zpi,Sabti:2021reh}.}
    \label{fig:moneyplot}
\end{figure}

In this section, we present the main results of our analysis. The dominant flavor component of neutrino flux in all of the experiments under consideration is muon neutrinos with a smaller contribution from electron neutrinos.  
Furthermore, in our scenario, the cross-section is identical for neutrinos and anti-neutrinos; thus we do not distinguish between incoming neutrinos or anti-neutrinos in our analysis. 

We note that the energy spectra of muon and electron neutrinos are relatively similar at the experiments we study, our qualitative conclusions apply to both scenarios. However, the electron neutrino flux is around two orders of magnitude lower than the muon neutrino flux at these experiments, therefore the constraints on the coupling of the new mediator to electron neutrinos is one order of magnitude weaker than the constraints on the coupling to muon neutrinos. In the following we describe the constraints on the coupling of the mediator to muon neutrinos in more detail.

In Fig.~\ref{fig:moneyplot} we show the 90\% C.L.\ exclusions and projected sensitivities in the $(m_\phi,\, c_\nu^\mu g_{\phi\gamma})$ and $(m_\phi,\, c_\nu^e g_{\phi\gamma})$ parameter space. Constraints from existing data -- MiniBooNE, MicroBooNE, T2K, and NOMAD -- are shown as solid lines and shaded regions. Projected sensitivities from experiments like ICARUS, SBND, and DUNE with 1 year and 10 years of running  are depicted with dashed lines. Additionally, we show the existing limits from XENONnT and cosmology, as discussed in the previous section. 

At low mediator masses, MiniBooNE---shown as the cyan shaded region in Fig.~\ref{fig:moneyplot}---provides the strongest existing constraint, excluding couplings down to $c_\nu^\mu g_{\phi\gamma} \lesssim 5.6 \times 10^{-6}\,{\rm GeV}^{-1}$ for $m_\phi \sim 1$ GeV and reaching  $c_\nu^\mu g_{\phi\gamma} \lesssim 1.4\times 10^{-7}\,{\rm GeV}^{-1}$ at MeV scale mass. In the heavy-mediator limit where $m_\phi$ can be integrated out, the MiniBooNE bound simplifies to $c_\nu^\mu (g_{\phi\gamma}\times{\rm GeV}) \lesssim 3.55 \times 10^{-6}\ (m_\phi^2/ \text{GeV}^2)$. Additional constraints from MicroBooNE and T2K, shown by the orange and blue lines respectively, provide complementary coverage in the sub-GeV to few-GeV mass range. At higher masses (\(m_\phi \gtrsim 4\)~GeV), the NOMAD experiment (pink) sets the leading bound.

We now contrast these existing limits with projected sensitivity from upcoming experiments. The  sensitivity of the SBND experiment, using the BNB flux and  assuming a total exposure of $6.6\times 10^{20}$ POT \cite{Bonesini:2022pwy}, is represented by the dashed black line in Fig.~\ref{fig:moneyplot}. It will probe couplings down to \(\sim 10^{-8}\,{\rm GeV}^{-1}\) in the low-mass region. The ICARUS experiment will receive neutrinos from both the NuMi and BNB beams. However, the projected sensitivity for the NuMI beam is weaker; therefore, we do not display it in Fig.~\ref{fig:moneyplot}. For the BNB beam, we assume an exposure of $3.8\times 10^{20}$ POT, corresponding to the existing amount of data they have collected already \cite{DiNoto:2024nwm}. The resulting forecasted constraint is shown in the black dashed line in Fig.~\ref{fig:moneyplot} and is slightly less stringent than the expected SBND result.

The forecasted results from the  DUNE liquid-argon near detector (LAr ND) are shown in purple. 
We present results for both 1 year and 10 years of running, assuming an annual exposure of 1.1$\times 10^{21}$ POT~\cite{DUNE:2020ypp}. DUNE is expected to achieve significant improvements: with one year of data, it could probe couplings down to \(c_\nu^\mu g_{\phi\gamma} \sim 10^{-9}\,{\rm GeV}^{-1}\), improving the existing bounds by at least two orders of magnitude for all mediator masses. With ten years of exposure, the sensitivity improves further by approximately a factor of 3, making DUNE the most sensitive probe for monophoton signal arising via neutrino polarizability. A similar study of  NC$1\gamma$ signature could be performed at the DUNE far detector. However, the neutrino flux at the far detector is significantly reduced, potentially resulting in weaker constraints. On the upside, at the far detector  electron and tau neutrinos that are produced by oscillations. This allows to probe couplings of $\phi$ to other neutrino flavors, offering complementary sensitivity to the model. We leave  the investigation of such possibility for future work.

In summary, MiniBooNE and NOMAD currently set the most stringent constraints across most of the parameter space, while SBND and DUNE will significantly extend the sensitivity to couplings up to three orders of magnitude over broad range of scalar masses, see tables~\ref{tab:exp_bounds_numu}, ~\ref{tab:exp_bounds_nue}  in Appendix \ref{sec:compilation} for a compilation of the results.

So far, we have discussed the constraints on the proposed framework from MiniBooNE and MicroBoone. Interestingly, these two experiments have also reported an excess of events in this channel. The low-energy electron like excess observed at MiniBooNE could originate from either electrons or photons in the final state. This excess has been extensively studied, but no conclusive explanation has yet been provided (see \cite{Abdullahi:2023ejc} for a recent review). Meanwhile, MicroBooNE has also reported an excess of events in the NC$1\gamma$ channel with a local significance of $2\sigma$. Since our model leads to monophoton signature that could potentially account for these anomalies, further investigation is warranted. 

 \begin{figure}[!t]
    \centering
     \includegraphics[width=0.9\linewidth]{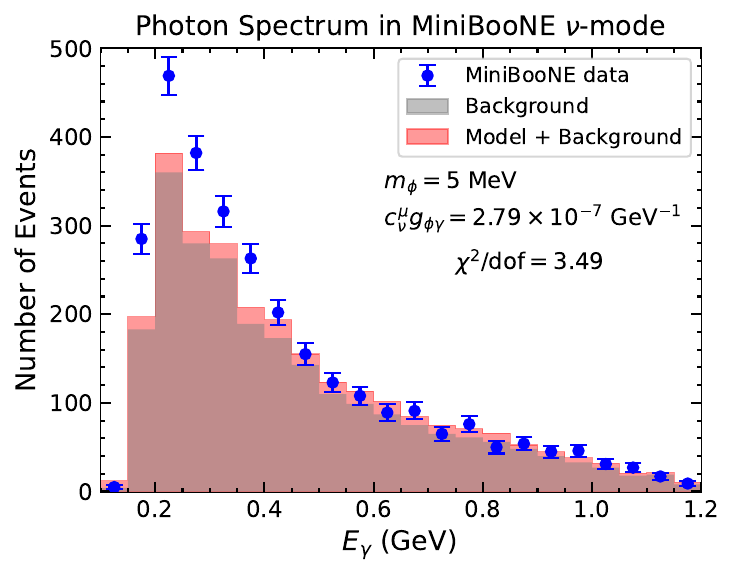} \\
     \includegraphics[width=0.9\linewidth]{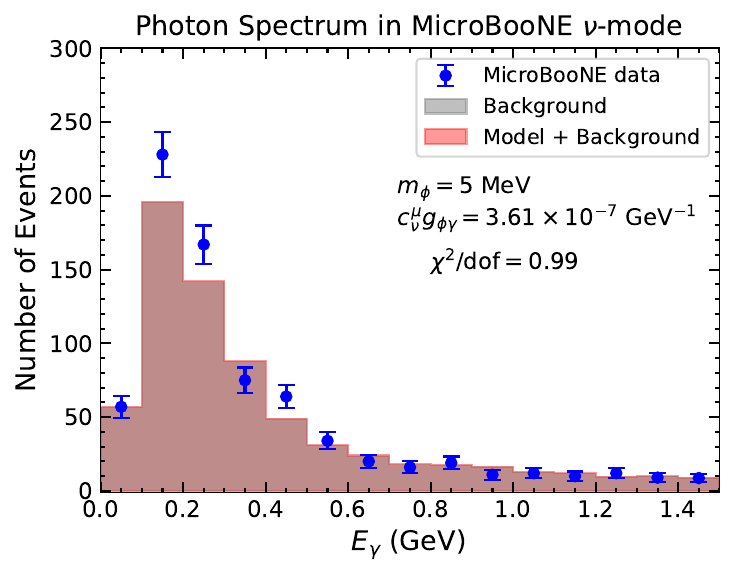}
    \caption{Photon energy spectrum from the MiniBooNE \cite{MiniBooNE:2020pnu}  (top) and MicroBooNE \cite{MicroBooNE:2015bmn} (bottom) neutrino mode data compared to our best fit model prediction added to the background. The best-fit model parameters respect other existing constraints.}
    \label{fig:miniboone_Eg}
\end{figure}

In Fig.~\ref{fig:miniboone_Eg} we show the energy spectrum for the neutrino mode at MiniBooNE for a best fit benchmark parameter point of $m_\phi=5$ MeV, $c_\nu^\mu g_{\phi\gamma}=2.79\times 10^{-7}~\text{GeV}^{-1}$ which avoids constraints from cosmology and MicroBooNE.  While our model provides a slightly better fit at higher photon energies than the background only hypothesis, it fails to reproduce the excess at low energies and can therefore only slightly alleviate the tension. We show the predicted angular distribution $\cos\theta_\gamma$ and anti-neutrino run in the Appendix~\ref{app:plots}. The model provides a better fit to the data in these modes compared to neutrino mode shown in Fig.~\ref{fig:miniboone_Eg} (upper plot). Our results agree with the conclusion in Ref.~\cite{Abdullahi:2023ejc}, namely that this model cannot fully explain the MiniBooNE excess. 

Turning now to the MicroBooNE data, we show the photon energy spectrum in Fig.~\ref{fig:miniboone_Eg} (lower plot) for the benchmark point of 
$m_\phi=5$ MeV, $c_\nu^\mu g_{\phi\gamma}=3.61\times 10^{-7}~\text{GeV}^{-1}$ which is allowed by constraints from cosmology and MiniBooNE at $3\sigma$. Note that the best-fit point from MicroBooNE lies in a region excluded by the MiniBooNE bound. For this choice of parameter, our model fits the data slightly better than the background-only hypothesis.

In conclusion, we conclude that the model  cannot fully explain the anomalous MiniBooNE and MicroBooNE data. This statement is equally true for the new mediator coupling  to either muon or electron neutrinos.

\section{Conclusions}

\label{sec:summary}
Neutrinos provide a promising window into physics beyond the Standard Model. As such testing their properties and comparing them to the Standard Model predictions can possibly unveil new physics. In this work, we focused on a less investigated electromagnetic property of neutrinos: {\it neutrino polarizability}, which couples two neutrinos to two photons. We have studied the current and near-term experimental sensitivity to neutrino polarizability via monophoton events at a variety of neutrino experiments. This effective dimension-7 interaction can be realized through the exchange of a pseudo-scalar mediator that couples to both neutrinos and photons. We have found no evidence of neutrino polarizability in available data sets. Among existing experiments, NOMAD currently sets the strongest bounds on the product of the pseudo-scalar couplings for mediator mass above a few GeV, while MiniBooNE sets the strongest constraints at lower masses. In the near-term SBND, ICARUS, and DUNE will offer a significant improvement in sensitivity by up to three orders of magnitude on the couplings over a wide range of mediator masses. These constraints from a monophoton signal are stronger than constraints from anomalous contributions to neutrino-electron scattering at dark matter experiments, and they are complementary to constraints from cosmology and astrophysical observables.

\section*{Acknowledgments}
We thank Michele Tammaro for  useful discussions.
JG acknowledges support by
the U.~S.~ Department of Energy Office of Science under award number DE-SC0025448.
JG and AT acknowledge useful discussions at the NEAT workshop at Colorado State University in May 2025 and thank the CERN Theory department for kind hospitality during the final stages of this work. IMS is supported by the U.S. Department of Energy Office of Science under award number DE-SC0020262.

\bibliography{main}
\appendix 
\section{Details on Form factors}
\label{sec:details_xsec}
In the coherent regime,
neutrinos scatter coherently off the entire nucleus
 since the momentum transfer is smaller than the momentum scale associated with the de Broglie radius of the nucleus. We use  the Helm form factor which is given by \cite{Dobrich:2015jyk}
\begin{equation}
F_{\rm Helm}\left(q^2\right) =\frac{3 j_1\left(\sqrt{q^2} R\right)}{\sqrt{q^2} R} \exp \left[-\frac{\left(\sqrt{q^2} s\right)^2}{2}\right]
\end{equation}
where $j_1$ is the spherical Bessel function of the first kind, and the parameter $R$ is defined as \cite{Lewin:1995rx}
\begin{equation}
R =\sqrt{\left(1.23 A^{1 / 3}-0.6\right)^2+\frac{7}{3} \pi^2 0.52^2-5 s^2},
\end{equation}
where we set $s = 0.9$ fm. The cross-section is then modified by a factor of $F_{\rm Helm}^2\left(q^2\right) \times Z^2$, where $Z$ is the atomic number of the target nucleus.

In the incoherent regime, where the neutrino scatters off the nucleon, we use a dipole form factor $F_{\rm dipole}$ to model the interaction:
\begin{equation}
    F_{\rm dipole} (q^2) = \frac{1}{1-q^2/m_s^2}, 
\end{equation}
with $m_s = 1.23 \pm 0.07$ GeV \cite{Kharzeev:2021qkd}. In the DIS regime ($q^2 > 1.8\ {\rm GeV}^2$), we compute the cross-sections using the parton distribution functions (PDFs) of the proton from the NNPDF23 PDF set.  

\section{Details on the \texorpdfstring{\pmb{$\chi^2$}}{chisq}  calculation}
\label{sec:details_analysis}
For our analysis, we vary the incoming neutrino energy $E_\nu$ from 0.05 GeV to $100$ GeV, depending on the experimental setup, and the scalar mass $m_\phi$ is scanned over the range $10^{-8} - 10^5$ GeV. For coherent scattering, we use carbon as the target nucleus for MiniBooNE, NOMAD, and T2K ND280 and liquid argon (LAr) for MicroBooNE, SBND, ICARUS, and DUNE.

To derive the bounds on the model parameters -- specifically the product of the couplings $(c_\nu g_{\phi\gamma})$ and the mass of the scalar field $m_\phi$ -- we perform a chi-squared analysis over the photon energy bins
\begin{equation}
\chi^2 = \sum_i \frac{\left(N_i^{\rm obs}-(N_{i}^{\rm bkg}+ N_i^{\mathrm{NP}})\right)^2}{\sigma_{i,{\rm stat}}^2 + \sigma_{i,{\rm syst}}^2} , 
\label{eq:chisq}
\end{equation}
where $N_i^{\rm obs}$, $N_i^{\rm bkg}$, and $N_i^{\rm NP}$ are the number of measured events, SM background events, and NP events. $\sigma_{i, {\rm stat} }^2 = N_i^{\rm obs}$ is the experimental statistical uncertainty and $\sigma_{i, {\rm syst} } = \frac{\sigma_{\rm syst}}{N^{obs}_{\rm tot}} N_i$ which we assume to be flat in energy.

For the future sensitivity projections, we adopt a Poissonian likelihood test statistic given by 
\begin{equation}
    \chi^2=2\sum_iT_i-D_i+D_i\log(D_i/T_i) \, ,
\end{equation}
We assume no excess events above the SM are detected, such that $D_i=0$ is the assumed data in each bin (unless otherwise stated) and $T_i$ is the predicted number of events from the BSM model. We impose the condition that the total number of predicted events remain below 2.7 at 90\% C.L.

\section{Experimental details}
\label{sec:exp_details}
In the following subsections, we review the relevant experimental setups in more detail and present comparisons between our theoretical predictions and the available data where applicable.

\subsection{MiniBooNE}
   
MiniBooNE was a short baseline experiment that uses a pure mineral-oil (CH$_2$) detector, which acts as both a target and a scintillator. It cannot distinguish an electron neutrino signature from a photon detection.  
The typical momentum transfer $q^2 \sim -2\ {\rm GeV}^2$, and the neutrino beam is primarily composed of $\nu_\mu + \bar{\nu}_\mu$ that peaks around $\sim 500$ MeV.  The total flux $\phi_{\rm tot} (E_\nu)$ in  [$\nu$/POT/GeV/$\text{cm}^2$] for both neutrino and antineutrino mode is given in Ref.~\cite{MiniBooNE:2008hfu}. The analyzed data sets uses $18.75 \times 10^{20}$ (neutrino) and $11.27 \times 10^{20}$ (antineutrino) POT. We use the photon detection efficiency $\epsilon (E_\gamma)$ from Ref.~\cite{MiniBooNE:2012maf}, with an energy threshold of $E_\gamma > 100\,\text{MeV}$~\cite{MiniBooNE:2020pnu, MiniBooNE:2018esg} and use $N_p = 2.8 \times 10^{32}$. We constrain  the proposed BSM model using a combined $\chi^2$ statistic: $ \chi^2 = \chi^2(\nu) + \chi^2(\bar{\nu})$, defined in eq.~\eqref{eq:chisq}, that accounts for observed events, background, and new physics signals.

\subsection{MicroBooNE}

MicroBooNE utilizes a liquid-argon time projection chamber (LArTPC) to search for single-photon signals. The neutrino flux is dominated by the Booster Neutrino Beam (BNB) 
and taken from Ref.~\cite{microbooneflux}. The detector has an active mass of 85 metric tons of liquid argon and we use $N_p = 1.28 \times 10^{30}$ target protons in our analysis. Current data has \(6.34 \times 10^{20}\) POT in neutrino mode with the projected future exposure of \(13.2\times10^{20}\) POT.  Photon efficiencies as a function of photon energy $E_\gamma$ and angle $\cos\theta$ are given in Ref.~\cite{MicroBooNE:2025ntu} (see supplemental materials). Since no explicit mapping from true photon energy to reconstructed shower energy is available publicly, we assume a one-to-one correspondence between them and minimum photon energy cut of  \(E_\gamma > 20\,\text{MeV}\) is applied~\cite{MicroBooNE:2022oim}.

\subsection{T2K}
The T2K near detector (ND280) has performed a dedicated search for neutral‐current single‐photon (NC1$\gamma$) events~\cite{T2K:2019odo} with the fiducial volume containing $N_p = 5.54 \times 10^{29}$ target nucleons~\cite{T2K:2019odo}. This analysis is based on a neutrino‐mode exposure of 
$5.738 \times 10^{20}\ \text{POT}$. We use Ref.~\cite{T2K:2012bge} for neutrino flux and use a selection efficiency of $1.9\%$ \cite{T2K:2019odo}. We assume the same efficiency factor for angular dependence $\cos\theta$ as well. We also adopt a conservative cut of \(E_\gamma > 50\,\text{MeV}\). Background‐subtracted event counts are obtained from Ref.~\cite{T2K:2019odo}. The total systematic uncertainty is taken to be \(26.83\%\) of the background, as quoted in table~1 of Ref.~\cite{T2K:2019odo}, and is applied uniformly across all energy bins.

\subsection{NOMAD}
The NOMAD experiment performed a search for single‐photon events using a drift-chamber  target with a fiducial mass of 2.7 tons, composed of carbon (64\%), oxygen (22\%), nitrogen (6\%), and hydrogen (5\%), corresponding to an effective atomic mass $A = 12.8$~\cite{NOMAD:2011gyy}. The incoming neutrino flux~\cite{NOMAD:2003owt} has an energy range of $E_\nu = 5-100 $ GeV with a total exposure of $5.1 \times 10^{19}$ POT. We use a constant single‐photon detection efficiency of 8\%~\cite{NOMAD:2011gyy} for both $E_\gamma$ and $\cos\theta$. We impose the photon-energy-angle cut $E_\gamma\,[\,1 - \cos\theta\,] \;\leq\; 0.05$, where $E_\gamma$ is the photon energy and $\theta$ is the polar angle relative to the beam direction. We also assume that all of our signal events pass the cut on PAN (defined in \cite{NOMAD:2011gyy})  which depends on the energy deposition of the neutral particles in the ECAL. 

\subsection{SBND/ICARUS}
SBND is located at 110 m and ICARUS is located at 600 m from the target and both use the same Booster Neutrino Beam (BNB) flux as MicroBooNE. We use the flux from Ref.~\cite{sbndflux} for SBND.
To get the BNB flux prediction at ICARUS we rescaled by factor of $(110/600)^2$. In addition to the BNB flux, ICARUS also sees NuMI neutrino beam, whose flux is taken from Ref.~\cite{icarusnumi}.
Neither SBND or ICARUS has performed a dedicated search for single photon. We therefore treat both detectors in a projected-sensitivity mode, adopting a constant photon-detection efficiency of 10\% (this is consistent with MicroBooNE efficiency). We also impose the same photon-energy cut used by MicroBooNE, $E_\gamma > 20$ MeV, to suppress the low energy background. 

SBND has a fiducial volume of 112 tons of LAr, which corresponds to $N_p = 1.69 \times 10^{30}$ protons, where as ICARUS has fiducial volume of 476 tons LAr, corresponding to $N_p = 7.17 \times 10^{30}$ protons. Current ICARUS running has delivered $3.8 \times 10^{20}$ POT on BNB and $3.42 \times 10^{20}$ POT for neutrino mode in FHC configuration, whereas for NuMI it delivered $2.28 \times 10^{20}$ POT for antineutrino mode in RHC \cite{DiNoto:2024nwm}. Although SBND has just begun taking data, no single-photon analysis exists, we therefore take $6.6\times10^{20}$ POT  and uniform 10\% efficiency across all bins, for simplicity. Slight change in POT have very little to no impact on our analysis.  

\subsection{DUNE}
DUNE will have excellent energy reconstruction and the ability to distinguish photos from electrons. We focus on the liquid-argon near detector (LAr ND) located 574 m from the target. It will have a 67 ton fiducial mass that corresponds to roughly $N_p = 1.01 \times 10^{30}$ protons. We use the flux and photon efficiency from DUNE TDR Ref.~\cite{DUNE:2020ypp}. We also impose a $E_\gamma > 20$ MeV cut and assume $1.1 \times 10^{21}$ POT per year and restrict the incoming neutrino energy to be $E_\nu = [0.5,10]$ GeV.

\section{Additional results}
\label{app:plots}
\begin{figure*}[!t]
    \centering
    \includegraphics[width=0.325\textwidth]{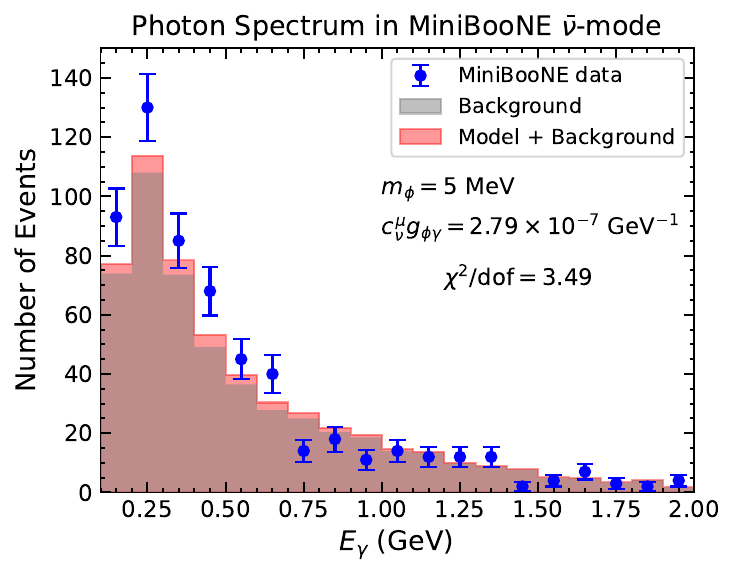}
    \includegraphics[width=0.325\textwidth]{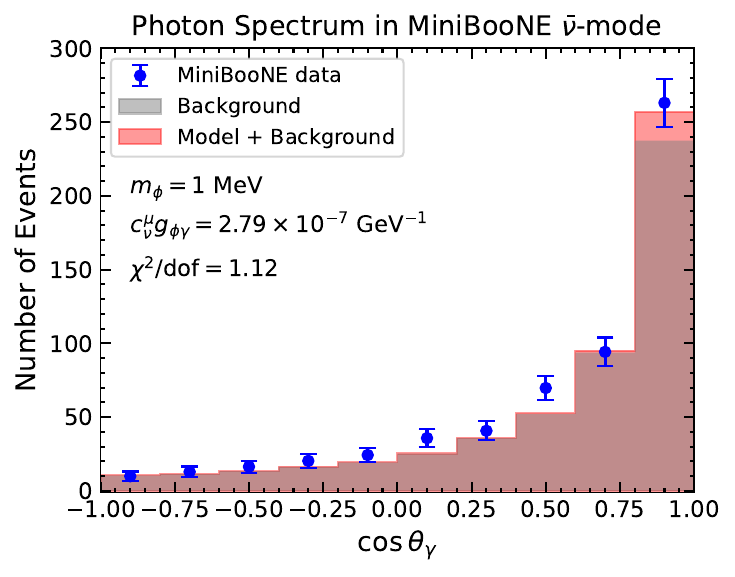}
    \includegraphics[width=0.325\textwidth]{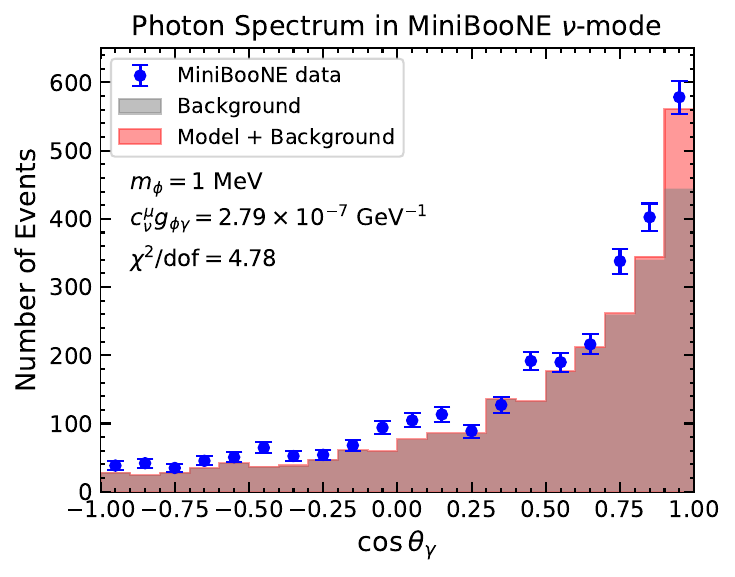}
    \caption{The photon energy and angular distribution spectrum for the MiniBooNE anti-neutrino run (left, middle) and the photon spectrum for the MiniBooNE neutrino run (right) \cite{MiniBooNE:2020pnu,MiniBooNE:2018esg}. We show the best fit model prediction added to the background. The best-fit model parameters respect other existing constraints.}
    \label{fig:mb_antinu}
\end{figure*}

We show the photon energy spectrum and angular distribution for the MiniBooNE anti-neutrino and neutrino run in Fig.~\ref{fig:mb_antinu}, see Fig.~\ref{fig:miniboone_Eg} for the energy spectrum in the neutrino run.

\section{Relevant other constraints}
\label{sec:constraints}
We summarize briefly the existing bounds on the parameter space of the model, namely $g_{\phi\gamma}$, $c_\nu$, and $m_\phi$. A comprehensive overview of these limits can be found in Ref.~\cite{Bansal:2022zpi}. 
\begin{enumerate}
    \item {\bf Cosmological constraints:} Cosmological observations impose one of the most stringent bounds on the parameter space for neutrino polarizability. First, the decay of $\phi$ into light relativistic particles, i.e., photons or neutrinos, after photon-neutrino decoupling shifts the effective number of neutrino species, $N_{\rm eff}$. Planck measurement require $N_{\rm eff} = 2.99_{-0.33}^{+0.34}$ at 98\% C.L.~\cite{Planck:2018vyg}.  Consequently, if $\phi$ decays predominantly to $\gamma \gamma$, the mass range $m_\phi = [1,10^3]$ keV is excluded. If $\phi$ only decays to neutrinos, $N_{\rm eff} = 3.57$. Additionally, during BBN, the thermal $\phi$ production will increase $N_{\rm eff}$. Requiring that $\phi$ not reach equilibrium at the photon-neutrino decoupling temperature yields constraint on $c_\nu$ \cite{Escudero:2019gvw}.
    Note that BBN constraints on $c_{\phi\gamma}$ are weaker than those obtained from the Planck data \cite{Depta:2020wmr}.
    
    The scalar field $\phi$ can induce a four-fermion operator $\nu\bar{\nu} \nu \bar\nu$ and lead to neutrino self interaction, which lead to $c_\nu < 0.28 (m_\phi /{\rm MeV})$ \cite{Lancaster:2017ksf} for $m_\phi \gg 100$ eV. For details including the case when $m_\phi \ll 100$ eV, see Ref.~\cite{Escudero:2019gvw}. In addition, neutrino decay can also put a constraint on the parameter space. Assuming $m_\phi \gg m_\nu$, two body decay of neutrino is not kinematically accessible. 
    Neutrino can, however, decay via three body final state $\nu_i \to \nu_j \gamma \gamma$. By requiring the neutrino lifetime is larger than the age of the Universe, the bound on the couplings read as $c_{\nu} (g_{\phi \gamma} \times {\rm GeV}) \lesssim 8\times 10^3 (0.1{\rm eV}/m_\nu)^{7/2}\ \left(m_\phi/\rm keV\right)^2$ \cite{Bansal:2022zpi}. 
Finally, if  $\phi$ interacts with neutrinos and photons it delays the process of neutrino decoupling \cite{Escudero:2018mvt,Serpico:2004nm,Sabti:2019mhn}. Constraints from CMB and BBN analyses place a lower limit on the mass of $\phi$, $m_\phi\gtrsim5$ MeV \cite{Sabti:2021reh}. This value changes slightly depending on the combined data sets.
  
    \item {\bf Stellar cooling:}  If the light scalar $\phi$ is produced in the stellar interior it can lead to excessive stellar cooling rates, which depends on the $\phi$ production rates as well as its decay length. The relevant $\phi$ production mechanisms are the Primakoff conversion ($\gamma \to \phi$), photon coalescence ($\gamma \gamma \to \phi$), and neutrino coalescence ($\nu \nu \to \phi$). Note that $\nu \nu \to \phi$ is not relevant for stellar objects with core temperature $T\sim {\cal O}$ keV scale, such as Horizontal Branch stars (HB), Red Giants (RG), and White Dwarves (WD), as it freely escapes these environments. Requiring $\phi$-cooling to remain below the experimental error on the measurement $|\epsilon_\phi| \leq 10\ {\rm erg}\ {\rm g}^{-1}\ {\rm s}^{-1}$~\cite{Raffelt:1987yb,Haft:1993jt} excludes $g_{\phi \gamma} \sim 10^{-11}\ {\rm GeV}^{-1}$ for $m_\phi < {\cal O}$ keV. If $m_\phi \gg {\cal O}$ keV, on-shell production is kinematically forbidden, and the cooling mechanism due to a plasmon-mediated Primakoff conversion $\gamma^*\gamma_L \to \nu \nu$ takes place. In that regime, one finds $ c_\nu (g_{\phi \gamma} \times {\rm GeV}) \lesssim 2.2 \times 10^{-3} \left(m_\phi/\rm MeV\right)^2$. 

    In a supernova with core temperature ${\cal O}$ (30 MeV), production mechanism via neutrino coalescence becomes important. The decay of $\phi \to \nu \nu$ will not contribute to the cooling rates, as neutrino produced inside the core are effectively trapped. By demanding that the total luminosity $L_\phi$ does not exceed the measured neutrino luminosity $L_\phi \lesssim L^\nu_{\rm SM} \sim 3 \times 10^{52}\ {\rm erg}\ {s}^{-1}$  lead to a very strong bound which we defer to the detail calculations done in Ref.~\cite{Lucente:2020whw,Fiorillo:2022cdq,Bansal:2022zpi} where the constraints can be read off.
    
    \item {\bf Other terrestrial constraints:} There are various other constraints which we list below without giving specific details:
    
    The three body decay $M^\pm \to \ell^{\pm}_\alpha \nu_\beta \phi$ can lead to strong constraint on the coupling $c_\nu$ if $m_\phi < m_M - m_{\ell_\alpha}$ \cite{Blinov:2019gcj}. Note that this constraint depend on the neutrino flavor. The relevant bound comes from $K^+ \to \ell^+ \nu \phi$, where $\ell = e, \mu$ (for $m_\phi \ll m_K$): $c_{\nu}^{e\alpha} \lesssim 4 \times 10^{-3}$ and $c_{\nu}^{\mu\alpha} \lesssim 1.5 \times 10^{-2}$. The decay of $\tau \to \ell \nu \bar{\nu} \phi$ also lead to $c_\nu \lesssim 0.3$ for light $\phi$.
    
   The emission of $\phi$ from a neutrino line in neutrinoless double beta decay experiment can put a strong constraint on $c_\nu^{ee}$ \cite{Blum:2018ljv}. The bound reads as $c_\nu^{ee} \lesssim 10^{-5}$ for $m_\phi \ll 2$ GeV and $c_\nu^{ee} \lesssim 10^{-4}$ for $m_\phi \sim 2$  GeV. 
    
   The field $\phi$ can be produced in electron and proton beam dump experiments via the Primakoff process. Its subsequent decay to a pair of photons can put a bound on the couplings $g_{\phi \gamma}$ \cite{Dolan:2017osp}. In our case, since $\phi$ has additional decay mode to a pair of neutrinos, the $\gamma \gamma$ signal is diluted. Thus one can recast the bound on $g_{\phi\gamma}$ from Ref.~\cite{Dolan:2017osp} by appropriately taking into account for the diphoton branching ratio. 
   
    Belle II’s search for $e^+e^-\to\gamma\,(\phi\to\gamma\gamma)$ gives the constraint on $g_{\phi\gamma}$ in the mass range $0.2 \lesssim m_\phi \lesssim 9.5$ GeV. For a fixed mass $m_\phi =1$ GeV, the coupling is constrained to be $g_{\phi\gamma} \lesssim 10^{-3}\,\mathrm{GeV}^{-1}$ at $m_\phi = 1$ GeV, assuming $\mathrm{Br}(\phi\to\gamma\gamma)=1$~\cite{Belle-II:2020jti}. 

\end{enumerate}

\section{Compilation of results}
\label{sec:compilation}
In tables~\ref{tab:exp_bounds_numu}, \ref{tab:exp_bounds_nue} we compile the constraints on our scenario for heavy mediators and $m_\phi=1$ GeV and $m_\phi=1$ MeV.

\begin{table}[!h]
\centering
\begin{tabular}{|l|c|c|c|}
\hline
Experiment & EFT bound & $m_\phi=$&$m_\phi=$\\
    &  $\left[m_\phi^2/\text{GeV}^2\right]$ &  1 GeV & 1 MeV \\
\hline \hline
MiniBooNE     & $\lesssim 3.55 \times 10^{-6}  $ & $5.6\times10^{-6}$ & $1.4\times10^{-7}$   \\
MicroBooNE &   $\lesssim 1.27 \times 10^{-5}  $ & $2.4\times10^{-5}$ & $4.9\times10^{-7}$   \\
T2K          &   $\lesssim 1.37 \times 10^{-4}  $ & $2.8\times10^{-4}$ & $5.2\times10^{-6}$    \\
NOMAD         &  $\lesssim 1.78 \times 10^{-7}  $ & $1.3\times10^{-5}$ & $2.2\times10^{-6}$  \\
SBND BNB     & $\lesssim 2.87 \times 10^{-7}  $ & $3.0\times10^{-7}$ & $3.5\times10^{-9}$   \\
ICARUS BNB    &  $\lesssim 7.60 \times 10^{-7}  $ & $8.0\times10^{-7}$ & $9.4\times10^{-9}$   \\
ICARUS NuMI   &  $\lesssim 9.4 \times 10^{-7}  $ & $3.8\times10^{-6}$ & $1.5\times10^{-7}$   \\
DUNE (1 year)     &  $\lesssim 1.61 \times 10^{-8}  $ & $2.3\times10^{-8}$ & $1.3\times10^{-9}$    \\
DUNE (10 year)   & $\lesssim 5.10 \times 10^{-9}  $ & $7.4\times10^{-9}$ & $4.0\times10^{-10}$    \\
\hline
\end{tabular}
\caption{Bounds and sensitivities from different neutrino experiments. Here the bounds are quoted for the product of couplings $c_{\nu}^\mu (g_{\phi \gamma} \times {\rm GeV})$ and the EFT bound is taken when $m_\phi^2 \gg q^2$, where $q$ is the typical momentum transfer in each experiment.  }
\label{tab:exp_bounds_numu}
\end{table}

\begin{table}[!t]
\centering
\begin{tabular}{|l|c|c|c|}
\hline
Experiment & EFT bound & $m_\phi=$&$m_\phi=$\\
    &  $\left[m_\phi^2/\text{GeV}^2\right]$ &  1 GeV & 1 MeV \\
\hline \hline
MiniBooNE     & $\lesssim 3.43 \times 10^{-5}  $ & $6.6\times10^{-5}$ & $1.9\times10^{-6}$   \\
MicroBooNE &   $\lesssim 1.04 \times 10^{-4}  $ & $3.1\times10^{-4}$ & $8.0\times10^{-6}$   \\
T2K          &   $\lesssim 7.33 \times 10^{-4}  $ & $1.6\times10^{-3}$ & $5.3\times10^{-5}$    \\
NOMAD         &  $\lesssim 2.68 \times 10^{-5}  $ & $1.7\times10^{-4}$ & $2.7\times10^{-5}$  \\
SBND BNB     & $\lesssim 1.38 \times 10^{-6}  $ & $2.8\times10^{-6}$ & $5.0\times10^{-8}$   \\
ICARUS BNB    &  $\lesssim 3.64 \times 10^{-6}  $ & $7.5\times10^{-6}$ & $1.3\times10^{-7}$   \\
ICARUS NuMI   &  $\lesssim 6.2 \times 10^{-6}  $ & $2.9\times10^{-5}$ & $9.4\times10^{-7}$   \\
DUNE (1 year)     &  $\lesssim 8.52 \times 10^{-8}  $ & $2.0\times10^{-7}$ & $1.2\times10^{-8}$    \\
DUNE (10 year)   & $\lesssim 2.70 \times 10^{-8}  $ & $6.3\times10^{-8}$ & $3.9\times10^{-9}$    \\
\hline
\end{tabular}
\caption{Bounds and sensitivities from different neutrino experiments. Here the bounds are quoted for the product of couplings $c_{\nu}^e (g_{\phi \gamma} \times {\rm GeV})$ and the EFT bound is taken when $m_\phi^2 \gg q^2$, where $q$ is the typical momentum transfer in each experiment.  }
\label{tab:exp_bounds_nue}
\end{table}

\end{document}